\documentclass{PoS}
\usepackage{graphicx}
\usepackage{epstopdf}
\newcommand{\lppr}{\stackrel{<}{\scriptstyle \sim}}

\title{Rapid VHE variability in blazars}

\ShortTitle{Rapid VHE variability}

\author{Francesca Volpe $^a$ and Frank M. Rieger $^{a,b}$ \\
\llap{$^a$} Max-Planck-Institut f\"ur Kernphysik, Saupfercheckweg 1, 69117 Heidelberg, Germany\\
\llap{$^b$} European Associated Laboratory for Gamma-Ray Astronomy, jointly supported by CNRS and MPG\\   
E-mail: \email{francesca.volpe@mpi-hd.mpg.de; frank.rieger@mpi-hd.mpg.de}}

\abstract{Active Galactic Nuclei (AGN) are known to show significant variability over a wide frequency 
range. We review observational results on the variability characteristics of blazars in the very high energy 
(VHE) domain, focusing on recent findings of rapid VHE variability and evidence for an underlying multiplicative 
driving process in PKS~2155-304. We explore a physical scenario where the variability is assumed to arise 
due to accretion disk fluctuations transmitted to the jet, and discuss its implications for the central powerhouse.}

\FullConference{25th Texas Symposium on Relativistic Astrophysics - TEXAS 2010\\
		December 06-10, 2010\\
		Heidelberg, Germany}

\begin{document}

\section{Introduction} \label{intro}

\begin{figure}[thb]
  \centering \includegraphics[width=8.8cm]{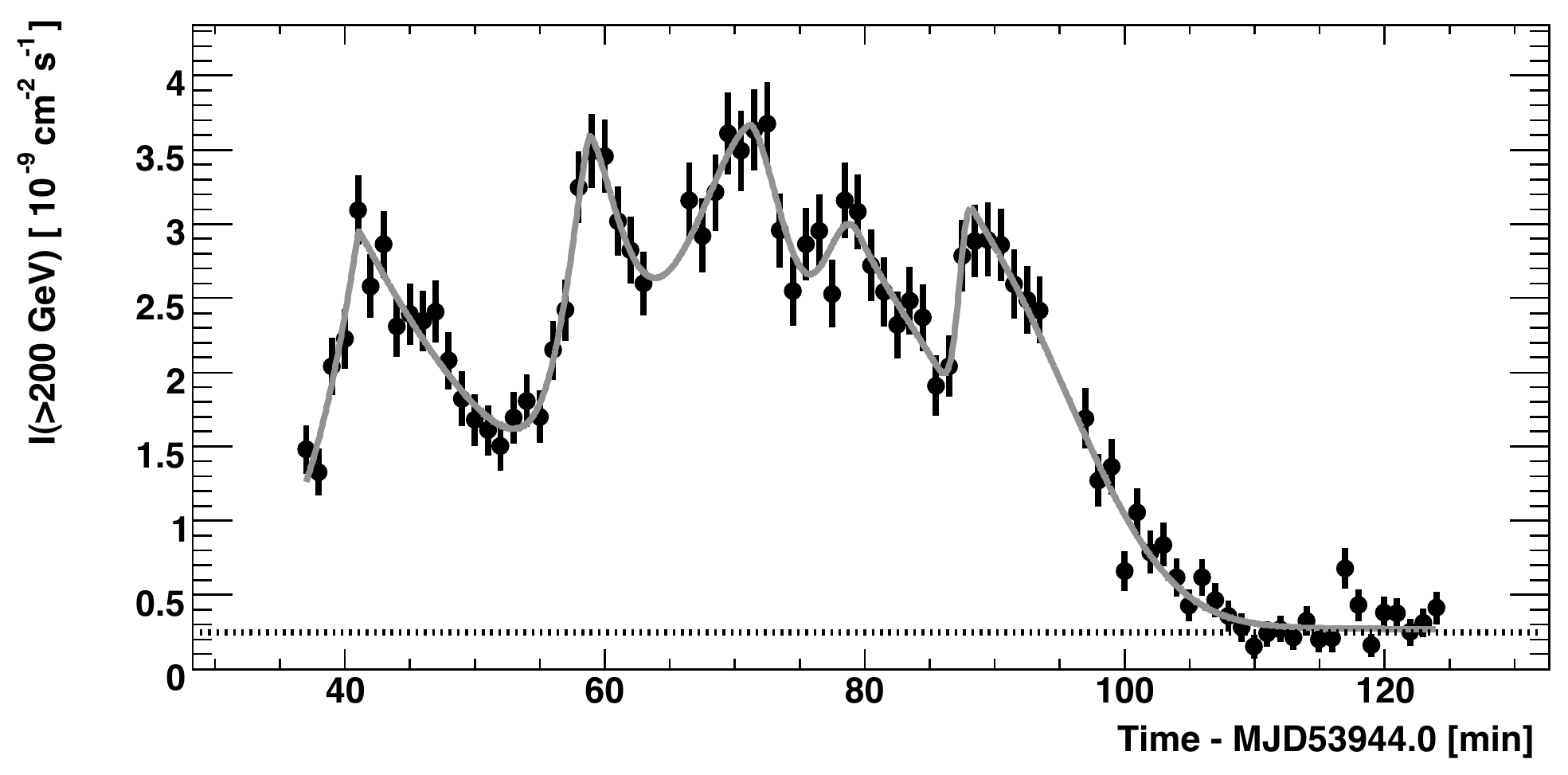}
  \centering \includegraphics[width=6cm]{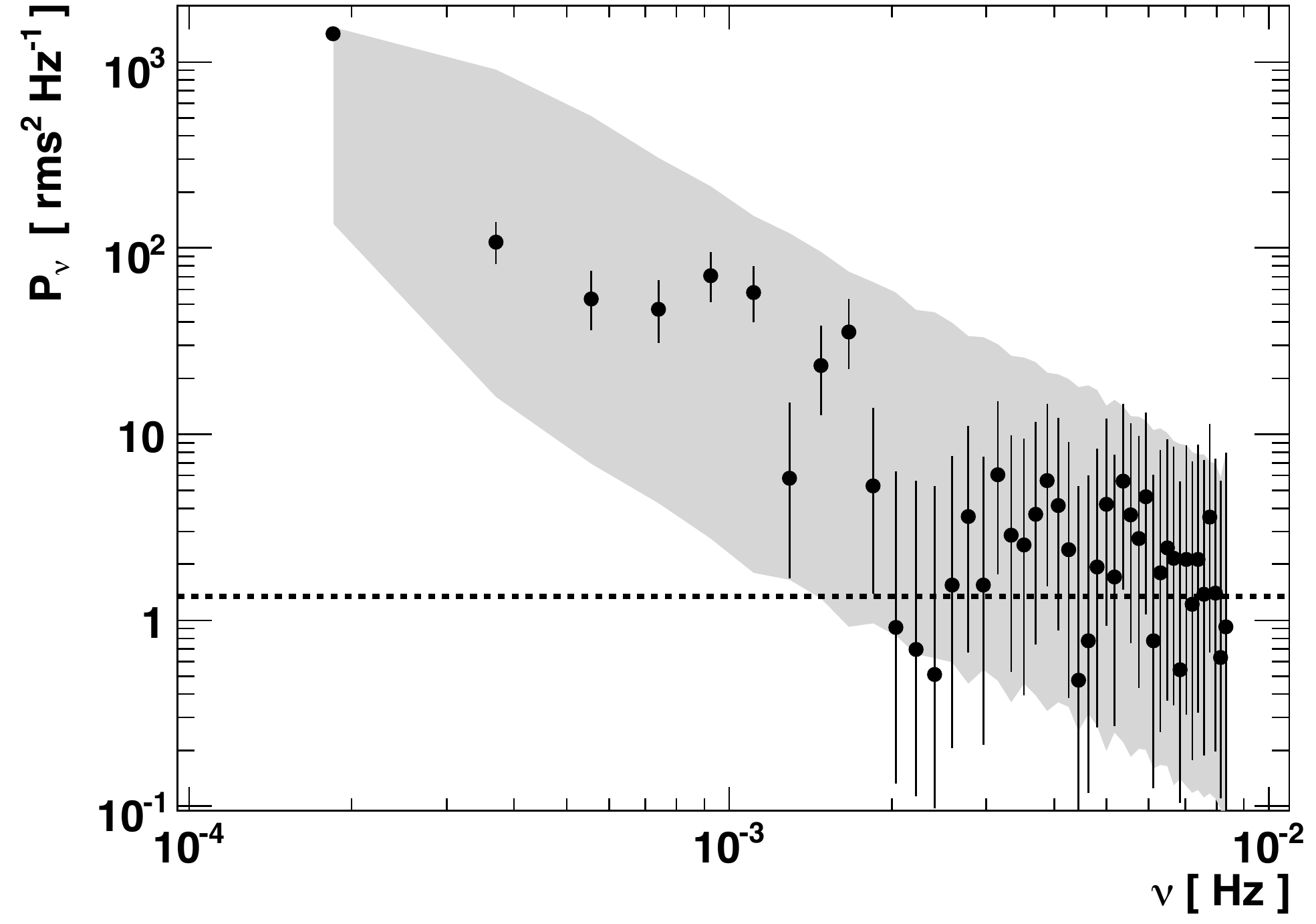}
  \caption{Left: H.E.S.S. integral flux above 200 GeV observed from PKS~2155-304 on MJD 53944 vs.
time. The data are binned in 1 minute intervals. Taken from \cite{HESSflare}. 
Right: The Fourier power spectrum of the light curve aside. The area corresponds to the 90\% CL for 
a power-law spectrum of index $-2$ \cite{HESSflare}.}  
  \label{fg1}
\end{figure}

Active Galactic Nuclei (AGN) show significant variability over a large range of timescales and with 
different amplitudes \cite{abdo10,gupta08,fan07}. At the time when this paper is written, despite the 
limited temporal coverage of Atmospheric Cherenkov Telescopes (ACTs) half of the AGN detected 
in the TeV domain by these experiments have shown variability. For the majority of them, variability 
timescales above one month have been found. In about a quarter of them there is clear evidence for 
short-term VHE variability on observed timescales of less than one day. The class of the high-frequency 
peaked BL-Lac objects currently reveals the most rapid VHE gamma-ray flux variability (observed
VHE variability timescales of a few minutes), as found by the H.E.S.S. and MAGIC experiments in 
PKS 2155-304 \cite{HESSflare} and Mkn 501 \cite{mkn501}, respectively. The latter findings suggest 
that one of the most constraining requirements on the jet kinematics and the high-energy emitting 
region may actually come from VHE variability studies.\\
Perhaps the most prominent example concerns the TeV blazar PKS 2155-304 at redshift of $z=0.116$.
In July 2006, the H.E.S.S. telescope array detected a dramatic VHE outburst in PKS~2155-304 
\cite{HESSflare}, with flux level varying between 1 and 15 Crab units and clear evidence for 
minute-timescale (t$_v \sim 200$ sec) VHE variability (see Fig. \ref{fg1}, left panel).\\ 
Apart from the extreme minimum variability timescale, Fourier analysis of the VHE light curve also 
revealed a red-noise-type PSD (power spectra density) with an exponent close to $\sim$ 2 within 
the frequency range $[$10$^{-4}$-10$^{-2}$Hz$]$ (see Fig. \ref{fg1} right panel). Similarly to results
in the X-ray domain, a correlation between (absolute) rms (root mean square) variability amplitude 
and mean VHE flux has been observed \cite{HESS_TEMP_VAR}. This is known to be characteristic 
of a non-linear, log-normal stochastic process where the relevant, normally distributed variable is
the logarithm of the flux $\log(X)$, and not just the flux $X$ itself \cite{uttley,mchardy}. The process
driving the VHE variability in PKS 2155-304 is therefore expected to satisfy a log-normal distribution.\\
In most currently-proposed scenarios for extreme short-term variability in blazars, this second finding 
is not discussed and this still leaves us with a challenge to coherently explain both facts. In the present
paper, we discuss a model (explained in more detail in \cite{rieger2}), where fluctuations in accretion 
disk rate feeding the jet are ultimately responsible for the observed PSD (Fig.~\ref{fluctuation_model}).

\section{On the log-normal VHE behavior in PKS~2155-304}\label{longormal}
\begin{figure}[!]
  \centering \includegraphics[width=12cm]{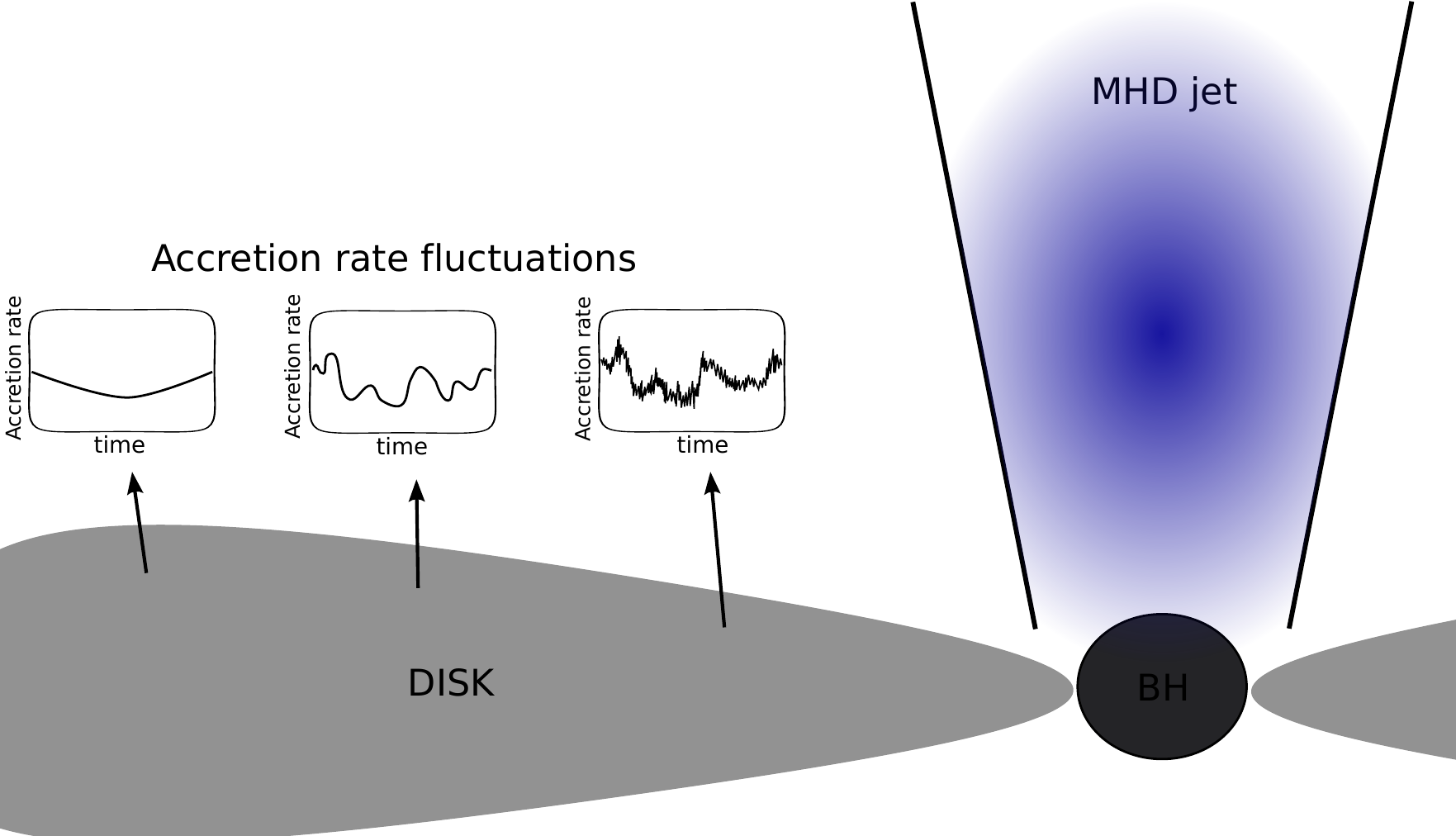}
\caption{Sketch of the considered model, where the VHE variability is driven by variations in the 
accretion rate: Small, independent fluctuations of the accretion rate on local viscous timescale 
$t_{\rm visc}(r)$ are assumed to occur over a range of disk radii $r\gg r_i$ large compared to 
the inner radius $r_i$ of the disk. If not damped, these fluctuations can propagate inwards and 
couple together to produce log-normal accretion rate variability close to $r_i$ of the flicker or 
red-noise-type. Any emission process linked to this region (e.g., plasma injected into the jet) 
may then eventually be modulated over a frequency interval ranging from the inverse accretion 
time near the outer to the one at the innermost disk radius, respectively.}\label{fluctuation_model}
\end{figure}

In principle, log-normal variability, as found in PKS 2155-304, can be considered as a statistical property 
of variable objects which is generated by a stationary random process by taking the exponential of a 
Gaussian time series. In such a case, the average fluxes is strongly correlated with the higher moments of 
the time series and thus with the rms. The correlation then signifies that the stochastic process driving this 
variability is multiplicative (not additive).\\
The log-normal VHE flux distribution in PKS 2155-304 can thus be thought of as the results of many 
multiplicative random effects as in a cascade. Variations on short timescales (determining the rms) will 
then obviously decrease in amplitude when the long timescales variations (determining the mean flux) 
decrease \cite{mchardy}.\\
In the context of accretion disk variability, Lyubarskii \cite{lyub} has shown that fluctuations of the 
disk parameters at some radius, occurring on local viscous timescale, can lead to log-normal-type
variations in the accretion rate at smaller radii that are of the flicker- or red-noise-type (cf. Fig.
\ref{fluctuation_model}).\\
%
\section{The origin of the VHE variability in PKS~2155-304}\label{OriginVar}
\begin{figure}[!]
  \centering \includegraphics[width=12cm]{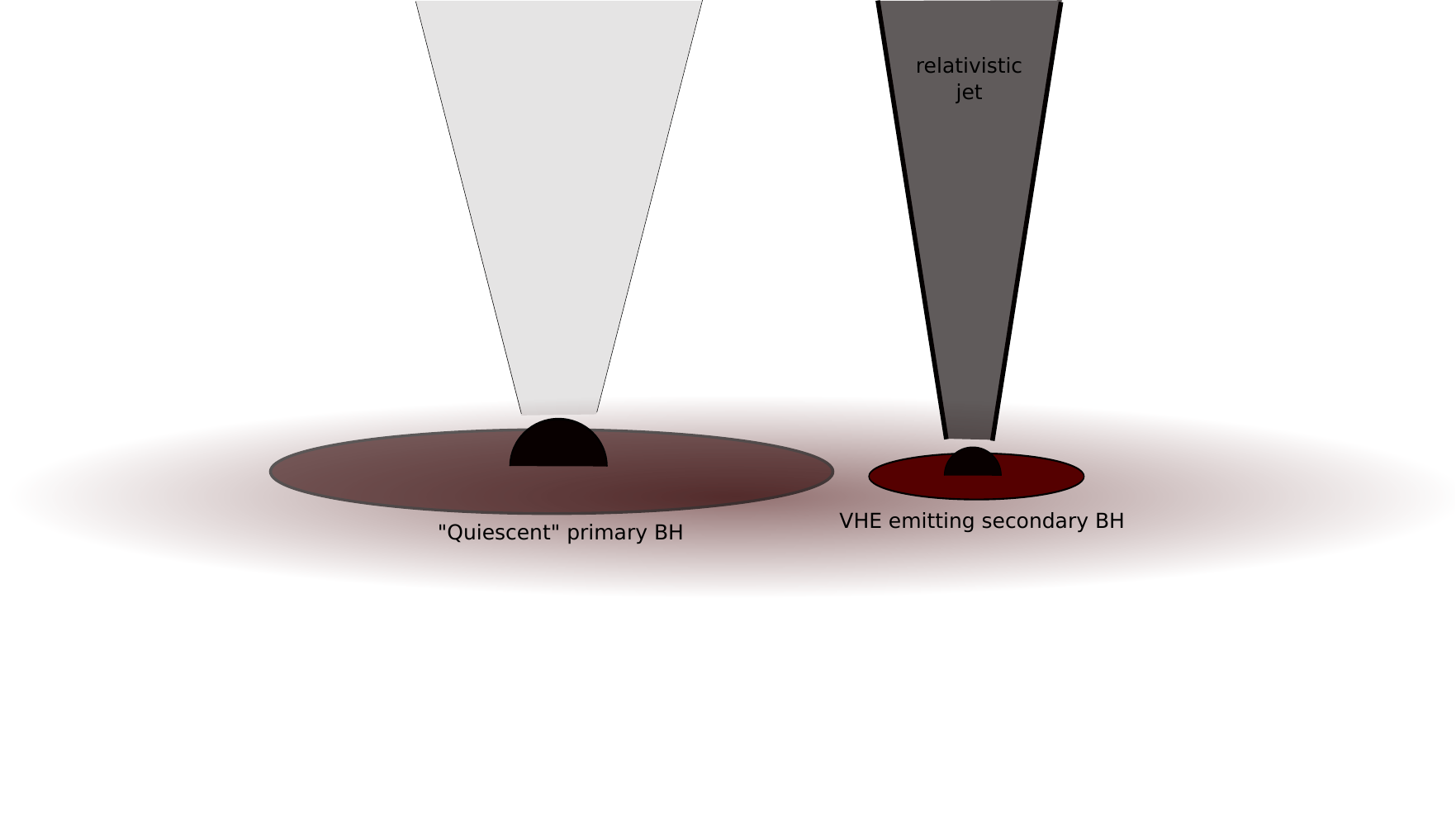}
\caption{Sketch of a possible binary black hole scenario for PKS 2155-304: Two black holes
orbit each other on quasi-coplanar orbits, with the binary surrounded by a circumbinary disk. 
Jet emission from the less massive, secondary black hole dominates the observed VHE radiation 
spectrum. Once the secondary becomes embedded in the outer disk around the primary, it starts 
clearing up an annular gap. Numerical simulations, e.g. \cite{artym,hayas07}, show however, that 
mass supply from the circumbinary disk to the central binary continues through tidal, time-dependent 
gas streams that penetrate the disk gap, periodically approaching and preferentially feeding the 
secondary (disk). The numerical results particularly evince a reversal of mass accretion rates 
can occur (i.e., $\dot{m_2} > \dot{m_1}$ despite $m_2 <m_1$), acting towards equal-mass binaries 
and suggesting that the secondary may become more luminous than the primary. If so, then applying 
variability constraints based on observed high emission properties can lead to a spurious 
identification of the secondary mass with the (total=primary+ secondary) central black hole mass.}  
  \label{binary_model}
\end{figure}

Suppose that these variations in the accretion rate are efficiently transmitted to the jet, resulting 
in red-noise fluctuations in the injection rate for Fermi-type particle acceleration. To successfully 
reproduce the variability characteristics in PKS 2155-304, one would then (at least) require the 
following:
\begin{enumerate}
\item First, the observed minimum variability timescale of $\sim 200$ sec, sets constraints on the 
size of the jet-emitting black hole dominating the VHE emission viz. $t_v \sim (1+z) r_g/c$, where 
$r_g=Gm/c^2$ is the gravitational radius of the black hole. The allowed maximum black hole 
mass would then be 
\begin{equation}\label{m_var}
m_v  \lppr 4 \times 10^7 \left(\frac{t_v}{200~\mathrm{sec}}\right)~M_{\odot}\,,
\end{equation} which is about an order of magnitude smaller than the anticipated central black hole 
mass inferred for PKS 2155-304 based on the known host galaxy luminosity relation \cite{rieger2}.
A rather small black hole mass also appears consistent with the X-ray variability properties observed
from PKS 2155-304 \cite{czern01,lach09}. Such apparently divergent mass estimates inferred from
high energy emission properties and host galaxy observations, could possibly indicate the presence 
of a close binary BH system \cite{rieger1,dermer1}, where the jet that dominates the high energy 
emission originates from the less massive (secondary) BH (see Fig.~\ref{binary_model}).
\item Secondly, we will only be able to observe rapid VHE emission (occurring on timescales as 
short as $200$ sec) with red noise-characteristics if these signatures do not get blurred by 
processes occurring on longer timescale within the source. As flux changes for an observer will 
always appear to be convolved and thus dominated by the longest timescale, this requires that 
the (observed) timescales for photons traveling across the radial width of the source and for the 
relevant radiative processes still remains smaller than $t_v$. As shown in \cite{rieger2}, this 
seems feasible in the case of PKS  2155-304.
\end{enumerate}
%
%
\section{A possible binary black hole scenario for PKS 2155-304}\label{BHmodel}
Suppose that the different mass estimates are indeed due to the presence of a close binary 
system (separation $d \ll 1$ pc), where the VHE emission is dominated by the jet from the less 
massive black hole. Such a binary system may be the outcome of the underlying hierarchical 
merging process shaping the host galaxy. There are a variety of reasons (e.g., observations 
of longterm periodicity) to take this to be of particular relevance for radio-loud AGN, e.g., see 
\cite{rieger3} for a review. Such a binary scenario for PKS 2155-304 may involve the following: 
\begin{enumerate}
\item While the observed short-term variability implies an upper limit on the secondary black 
hole mass, a lower limit can be derived based on the observed VHE luminosity during the 
outburst, i.e., the secondary mass $m_2$ cannot be too small if one also wishes to account 
for the required jet power. The analysis suggests that $m_2 \sim 10^7 M_{\odot}$ \cite{rieger2}.
\item There are indications for an optical longterm periodicity of $\sim (4-7)$ yr in PKS 2155-304
\cite{fan}. If true, this may fit nicely into a binary framework and suggest an upper limit on the 
intrinsic Keplerian orbital period of the binary of
\begin{equation}\label{kepler}
 P_k \leq  \frac{2}{(1+z)}\,P_{\rm obs}^{\rm opt} 
                \simeq 13 \left(\frac{P_{\rm obs}^{\rm opt}}{7~\mathrm{yr}}\right)~\mathrm{yr}\,,
\end{equation} if one assumes e.g. that the observed optical longterm periodicity is caused by 
the secondary crossing the disk around the primary twice per orbital period, or similarly, that
time-dependent gas streams are periodically feeding the central binary \cite{artym}.
\item The residual life-time of the binary system due to gravitational wave emission (assuming
quasi-circular orbits) is then expected to be below $\sim 10^8$ yrs \cite{rieger2}, providing
phenomenological support to the notion that supermassive binary black hole systems are 
able to coalescence within a Hubble time.
\item If the jet launched from the black hole is wrapped by a sufficiently strong magnetic field, 
its overall path can become curved (at least on small, sub-VLBA scales) due to the orbital 
motion of the supporting black hole. Then, the need for high minimum-bulk-Lorentz-factors of 
the outflow (as inferred from VHE SED modelling) might be somewhat relaxed because the 
effective Doppler factor becomes time-dependent \cite{rieger2}. This (and the possibility that
the primary may also produce a slower, radio-emitting wind) could help to explain why on 
larger (radio VLBA) scales only modest Doppler boosting appears to be present \cite{piner04}.
\item High-resolution VLBI images for PKS~2155-304 indicate strong jet bending within the 
inner milli-arcsecond (i.e., on parsec-scales) \cite{piner10}. This could possibly be caused
by precessional and/or orbital driving in a binary black hole system \cite{zensus97}.
\end{enumerate}

\section{Conclusions}\label{Conclusion}
The observed extreme VHE variability characteristics of PKS~2155-304 provide strong 
constraints on the physical parameters of its engine. We have suggested that the putative 
presence of a close supermassive binary binary system could allow to, e.g.,
\begin{itemize} 
\item reconcile central mass estimates based on host galaxy observations (indicative of 
the total primary and secondary mass) with those based on VHE gamma-ray variability 
(possibly only indicative of the jet-emitting secondary),
\item account for the observed log-normal variability characteristics via accretion disk 
fluctuations,
\item relax constraints on the jet flow velocity.
\end{itemize} 
Obviously, an increased instrumental sensitivity in the TeV domain (with a CTA-type 
instrument) that may allow us to search for even faster variability and, complementary, 
an advanced QPO analysis in the optical could be particularly valuable to assess the 
plausibility of such a scenario for PKS 2155-304.


%

\begin{thebibliography}{99}
%
\bibitem{abdo10} A.A. Abdo et al. (Fermi Collaboration) 2010, {\em ApJ}, {\bf 722},  520
\bibitem{gupta08} A.C. Gupta et al. 2008, {\em AJ}, {\bf 135}, 1384
\bibitem{fan07} J.H. Fan et al.\ 2007, {\em A\&A}, {\bf 462}, 547 
\bibitem{HESSflare} F. Aharonian et al. (H.E.S.S. Collaboration) 2007, ApJ, {\bf 664},  L71
\bibitem{mkn501} J. Albert et al. (MAGIC Collaboration) 2007, {\em ApJ}, {\bf 664}, 71
\bibitem{HESS_TEMP_VAR} A. Abramowski et al. (H.E.S.S. Collaboration) 2010, {\em A\&A}, {\bf A83}, 16 
 \bibitem{uttley} P. Uttley et al. 2005, {\em MNRAS}, {\bf 359}, 345
\bibitem{mchardy} I. McHardy 2008, {\em PoS(BLAZARS2008)}, {\bf 14}  
\bibitem{rieger2} F.M. Rieger \& F. Volpe  2010, {\em A\&A}, {\bf 520}, A23 
\bibitem{lyub} Y.E. Lyubarskii 1997, {\em MNRAS}, {\bf 292}, 679 
\bibitem{czern01} B. Czerny et al. 2001, {\em MNRAS}, {\bf 325}, 865 
\bibitem{lach09} P. Lachowicz et al. 2009, {\em A\&A}, {\bf 506}, L17
\bibitem{rieger1} F.M. Rieger \& K. Mannheim 2003, {\em A\&A}, {\bf 397}, 121 
\bibitem{dermer1} C. Dermer, J. Finke, G. Menon 2008, {\em PoS}(Blazar2008) 019
\bibitem{rieger3} F.M. Rieger 2007, {\em Ap\&SS} {\bf 309}, 1
\bibitem{artym} P. Artymowicz \& S.H. Lubow 1996, {\em ApJ}, {\bf 467}, L77 
\bibitem{hayas07} K. Hayasaki, S. Mineshige, \& H. Sudou 2007, {\em PASJ} {\bf 59}, 427
 \bibitem{fan} J.H. Fan \& R.G. Lin 2000, {\em A\&A}, {\bf 355}, 880 
\bibitem{piner04} B.G. Piner \& P.G. Edwards 2004, {\em ApJ}, {\bf 600}, 115
\bibitem{piner10} B.G. Piner, N. Pant \&  P.G. Edwards 2010, {\em ApJ}, {\bf 723}, 1150 
\bibitem{zensus97} A. Zensus 1997, {\em ARA\&A}, {\bf 35}, 607
%
\end{thebibliography}
\end{document}